\documentclass[aps,prb,reprint]{revtex4-1}
\usepackage{amsmath,amssymb,graphicx,xcolor,hyperref}
\usepackage[utf8]{inputenc}
\usepackage[english]{babel}  % LaTeX for English
\begin{document}

\title{Vortices and composite order in $\mathrm{SU}(N)$ theories coupled to Abelian gauge field}

\author{Daniel Weston}
\affiliation{Department of Physics, KTH Royal Institute of Technology, SE-106 91 Stockholm, Sweden}
\author{Egor Babaev}
\affiliation{Department of Physics, KTH Royal Institute of Technology, SE-106 91 Stockholm, Sweden}
\date{2019-08-28}

\begin{abstract}
We consider $\mathrm{SU}(N)$-symmetric Ginzburg-Landau models coupled to non-compact Abelian gauge field focusing on the case $N > 2$ at finite temperature. We show that, at least for sufficiently large gauge-field coupling constants, these models have two phase transitions. The intermediate phase between the symmetric and low-temperature phases is a state with composite neutral order and no Meissner effect. In this neutral phase the system spontaneously breaks only the symmetry associated with phase differences and density differences between components. For $N > 2$, in contrast to the $SU(2)$ case, the neutral state cannot be mapped onto an $\mathrm{O}(M)$ model. We term this state ${\mathbb{C}{P}}^{N-1}$-neutral phase. We also show that while $\mathrm{SU}(N)$-symmetric Ginzburg-Landau models are not superconductors or superfluids in the usual sense, their state in external field at sufficiently low temperature is a vortex lattice.
\end{abstract}

\maketitle
\section{Introduction}
\newcommand{\cp}{{\mathbb{C}{P}}^{N-1}}

Phase transitions in $\mathrm{U}(1)$ Abelian gauge theories, such as superconductors, is an old problem which is fairly well understood in the single-component case.\cite{Dasgupta1981,Peskin1978} The phase diagrams of multicomponent superconductors beyond mean-field approximation are more complex and less studied. Multicomponent gauge theories with Abelian gauge fields appear in the context of multicomponent electronic and nuclear superconductivity (see e.g.\ Refs.~\onlinecite{babaev2004superconductor, jones2006type, wang2017topological, wang2016topological}), and as effective field theories for other quantum systems.\cite{Senthil2004, Kuklov2006, Motrunich2008, Kuklov2008, Chen2013} The principal difference is that multicomponent models feature additional phases in which there is order only in various products of the complex fields, whilst individual phases are disordered.

The most investigated cases are two- and three-dimensional $U(1)\times U(1)$ superconductors described by two complex fields $\psi_{1,2}$. They  have, under certain conditions, a phase transition to a ``paired" state where order remains only in the products of the fields such as  $\langle\psi_1\psi_2\rangle \ne 0$ or $\langle \psi_1 \psi_2^* \rangle \ne 0$ while individual phases are disordered: $\langle \psi_i \rangle = 0$.\cite{babaev201547phase, babaev2004superconductor, Kuklov2006, Smiseth2005, Smorgrav2005b, Herland2010} The new states represent (i) neutral superfluids, where only counter-flow of charged components are allowed (termed metallic superfluid) and (ii) charge-4e superconductors. Analogous states of composite order occur for different microscopic reasons in strongly interacting superfluid mixtures on a lattice \cite{kuklov2004superfluid, kuklov2004commensurate, PhysRevLett.90.100401} and in pair-density-wave superconductors.\cite{berg2009charge, agterberg2008dislocations, radzihovsky2009quantum}

The origin of these transitions and additional phases is the fact that in these systems the composite topological defects, i.e.\ bound states of vortices in different components, have lower energy than the simplest vortices in the individual fields $\psi_{1,2}$. The entropically driven proliferation of composite vortices disorders the phases of the individual fields but does not restore the symmetry completely: the system retains order in products of the fields up to a higher temperature where the elementary toplogical defects proliferate (for a discussion of the principle see Ref.~\onlinecite{Svistunov2015}).

Besides representing new types of states, paired phases have attracted interest in models connected with the discussion of the order of the phase transitions in the so-called deconfined quantum criticality problem.\cite{Kuklov2006, Motrunich2008, Kuklov2008, Chen2013, Thomas1978} Phases with composite order have also been found in $\mathrm{SU}(2)$ models. \cite{Kuklov2008b, Motrunich2008, Kuklov2008, Herland2013} A theory for an $\mathrm{SU}(2)$ double of complex fields coupled to Abelian gauge field can be thought of as a model with $\mathrm{U}(1)_\mathrm{local} \times \mathrm{O}(3)_\mathrm{global}$ symmetry.\cite{banerjee2012atomic, zohar2015quantum}

The situation is more subtle in the $\mathrm{SU}(2)$ case than in the $\mathrm{U}(1)\times \mathrm{U}(1)$ case since it is believed that there are no energetically stable composite vortex-like topological defects in type-2 $\mathrm{SU}(2)$ gauge theories.\cite{achucarro2000semilocal} Nonetheless, for sufficiently strong coupling constant it has been shown that such $\mathrm{SU}(2)$ models have two transitions: at lower temperature the Meissner effect disappears but the system retains broken global $\mathrm{O}(3)$ symmetry which is restored only at a higher temperature. Currently there is no known duality argument relating phase transitions to statistical mechanics of topological defects in $\mathrm{SU}(N)$ cases of the kind that exist for $\mathrm{U}(1)$-systems.\cite{Dasgupta1981,Peskin1978} However, it should be noted that even the lack of energetic stability of the composite vortices in the ground state does not necessarily exclude entropically excited such vortices driving the phase transitions. In that respect an instructive example is the magnetic response of $\mathrm{SU}(2)$ gauge theory:\cite{Garaud2014} while strictly speaking $\mathrm{SU}(2)$ gauge theory is not a superconductor [i.e.\ does not have a conserved $\mathrm{U}(1)$ topological invariant \cite{svistunov2015superfluid}] and individual vortices carrying one flux quantum are not stable in the type 2 regime,\cite{achucarro2000semilocal} nonetheless in an external filed the system forms a vortex lattice \cite{Garaud2014} stabilized by intervortex forces. Likewise vortices can exist under certain conditions in neutral $\mathrm{SU}(2)$systems.\cite{Galteland2015}

The above raises the question of what the properties are of topological excitations and phase diagrams in $\mathrm{SU}(N)$ gauge theories with $N > 2$. Such higher-$N$ theories have attracted recent interest in the context of effective field theory for the quantium J-Q Heisenberg model, \cite{harada2013possibility, kaul2011quantum, kaul2012lattice} which catalyzed renewed interest in $\mathrm{SU}(N)$ gauge theories \cite{ihrig2019abelian, Fejos2017, gorbenko2018walking, nogueira2013deconfined}. Another motivation to study these states is the prospect of the creation of artificial gauage fields in ultracold atoms, which can open up new possibilities of studying these generalizations of superconducting states.

The paper is structured as follows: First we present the models that we consider and the Monte Carlo simulation methods that we use. Thereafter we present our work on phase diagrams, and finally we present our work on response to external magnetic field.

\section{Models with fixed total density}

We consider $\mathrm{SU}(N)$-symmetric Ginzburg-Landau models in three spatial dimensions, given by the Hamiltonian density
\begin{equation}
  h = \tfrac{1}{2} \sum_i |(\nabla + \mathrm{i}q\mathbf{A}) \psi_i|^2 +
      \tfrac{1}{2}(\nabla \times \mathbf{A})^2.
  \label{contf}
\end{equation}
Here $\mathbf{A}$ is the magnetic vector potential and the $\psi_i = |\psi_i| \mathrm{e}^{\mathrm{i}\phi_i}$ are matter fields corresponding to the superconducting components. The amplitudes are subjected to the constraint that the total superconducting density $\sum_i |\psi_i|^2 = 1$, but are otherwise allowed to fluctuate.

In order to elucidate the properties of the model (\ref{contf}) we note that it can be rewritten as 
\begin{multline}
  h = \frac{1}{2} \mathbf{j}^2
  + \tfrac{1}{2}(\nabla \times \mathbf{A})^2
         \\
      + \sum_{i,j>i} |\psi_i|^2 |\psi_j|^2 (\nabla \phi_{ij})^2 + \tfrac{1}{2} \sum_i \left( \nabla |\psi_i| \right)^2,
  \label{modes}
\end{multline}
where $\phi_{ij} = \phi_j - \phi_i$ and $\mathbf{j}$ is the density of charged supercurrent:
\begin{equation}
  \mathbf{j} = \sum_i |\psi_i|^2 (\nabla \phi_i + q \mathbf{A}).
  \label{current}
\end{equation}
Here the first line gives the the terms associated with electrically charged currents and magnetic field energy. On the second line we list those terms that give the energy from electrically neutral currents consisting of counterflows of charged components and gradients of relative density.

Note that, quite generically, when all components have non-zero density, only the vortices that have winding in the phase of each component (composite vortices) will have finite energy per unit length \cite{Babaev2002b}, as do vortices in ordinary single-component superconductors. These are therefore the energetically cheapest vortices that can be thermally excited. However, if  such objects proliferate, they cannot restore symmetries associated with the phase and density differences between components. Below we investigate numerically the possibility of this scenario.

\section{Monte Carlo simulation methods}

We discretize the model (\ref{contf}) on a three-dimensional simple cubic lattice with $L^3$ sites and lattice constant $a = 1$. The discretized model is given by the Hamiltonian density
\begin{equation}
  h = \tfrac{1}{2} \sum_{k<l} F_{kl}^2
      - \sum_{i,k} |\psi_i(\mathbf{r})|
        |\psi_i(\mathbf{r}+\mathbf{k})| \cos \chi_{i,k}(\mathbf{r})
  \label{lattf}
\end{equation}
where
\begin{equation}
  F_{kl} = A_k(\mathbf{r}) + A_l(\mathbf{r} + \mathbf{k})
           - A_k(\mathbf{r} + \mathbf{l}) - A_l(\mathbf{r})
\end{equation}
is a lattice curl,
\begin{equation}
  \chi_{i,k}(\mathbf{r}) =
    \phi_i(\mathbf{r}+\mathbf{k}) - \phi_i(\mathbf{r}) + q A_k(\mathbf{r})
\end{equation}
is a gauge-invariant phase difference, $k$ and $l$ signify coordinate directions, and $\mathbf{k}$ is a vector pointing from a lattice site to the next site in the $k$-direction. Periodic boundary conditions are applied in all three spatial directions. The thermal probability distribution for states of the system at inverse temperature $\beta$ is given by the Boltzmann weight
\begin{equation}
  \mathrm{e}^{-\beta H}, \quad H = \sum_{\mathbf{r}} h(\mathbf{r}).
\end{equation}
We generate representative samples from these thermal distributions using Monte Carlo simulation.

The simulations are performed using the Metropolis-Hastings algorithm with local updates of each of the degrees of freedom. The sizes of the local updates of the vector potential are adjusted during the initial part of the equilibration in order to make the acceptance probability $0.5$. Each phase is updated separately by simply selecting a new value for the phase with uniform distribution on the range from $-\pi$ to $\pi$. The amplitude degrees of freedom on a given site are updated together by randomly choosing a new density distribution. Care must be taken to use the correct measure when updating the amplitudes subject to the constraint of fixed total density.

For the determination of phase diagrams, we also use parallel-tempering swaps between systems with neighboring temperatures; typically one set of swaps is proposed every $24$ sweeps. The simulated temperatures are adjusted within a fixed interval in order to make the acceptance ratios for parallel-tempering swaps equal for all pairs of neighboring temperatures. We use reweighting between simulated temperatures in order to improve our data (more on this below). For the determination of low-temperature vortex configurations we use simulated annealing.

Equilibration is checked by comparing results obtained using the first and second halves of the data gathered after equilibration, and by comparing inverse-temperature derivatives obtained using finite differences and statistical estimators. Errors are determined by bootstrapping, and error bars correspond to one standard error.

\section{Phase diagrams}

In this section we consider the phase diagrams, in terms of electric charge $q$ and inverse temperature $\beta$, for the cases of two, three and four components. These phase diagrams involve two types of order: superconducting order and $\cp$ order. We first describe the methods we use to locate phase transitions of each type, and then present and discuss our results on phase diagrams.

\subsection{Locating superconducting transitions}

While systems that do not have local $\mathrm{U}(1)$ symmetry are not
superconductors in the usual sense [i.e.\ do not have a conserved $\mathrm{U}(1)$ topological invariant \cite{Svistunov2015}], they can exhibit magnetic field screening (the Meissner effect). Correspondingly we define a superconducting phase transition as a transition where the Meissner effect disappears. In order to locate superconducting transitions, we use the dual stiffness\cite{Motrunich2008, Herland2013, Carlstrom2015}
\begin{equation}
  \rho^\mu(\mathbf{q}) =
    \left\langle \frac{\left| \sum_{\mathbf{r},\nu,\lambda}
      \epsilon_{\mu\nu\lambda} \Delta_\nu A_\lambda(\mathbf{r})
      \mathrm{e}^{\mathrm{i}\mathbf{q} \cdot \mathbf{r}} \right|^2}
      {(2\pi)^2 L^3} \right\rangle,
\end{equation}
where $\epsilon_{\mu\nu\lambda}$ is the Levi-Civita symbol, $\Delta_\nu$ is a difference operator and $\langle \cdot \rangle$ is a thermal expectation value. Specifically, we measure the dual stiffness in the $z$-direction evaluated at the smallest relevant wave vector in the $x$-direction $\mathbf{q}_\mathrm{min}^x = (2\pi/L, 0, 0)$, i.e.\ $\rho^z(\mathbf{q}_\mathrm{min}^x)$; we denote this quantity simply as $\rho$. The quantity $\rho$ will (in the thermodynamic limit) be zero in the Meissner state in which fluctuations of the magnetic field are suppressed, and non-zero in the normal phase. In the sense that it is zero in the low-temperature phase and non-zero in the high-temperature phase, it is thus a dual order parameter.

The quantity $\rho$ is expected to scale as the inverse system size $1/L$ at the critical point of a continuous superconducting transition. Consequently, $L\rho$ is a universal quantity, the finite-size crossings of which (extrapolated to the thermodynamic limit) give the critical temperature of a superconducting transition.

\subsection{Locating neutral transitions}

The model has a neutral sector that we call $\cp$ neutral sector. Again, because there is no conserved $\mathrm{U}(1)$ topological invariant the order in that sector is not strictly speaking superfluid, in contrast to the composite order in $\mathrm{U}(1)^N$ models.\cite{babaev201547phase, Svistunov2015} In order to locate neutral transitions we use the helicity modulus, which measures the free-energy cost of an infinitesimal phase twist.

Consider imposing a twist in a certain linear combination $\sum_i a_i \phi_i$ of the phases, i.e.\ replacing the the phase $\phi_i$ by
\begin{equation}
  \phi_i'(\mathbf{r}) = \phi_i(\mathbf{r}) -
                        a_i\,\boldsymbol{\delta}\cdot\mathbf{r}.
\end{equation}
The helicity modulus for this phase combination is then
\begin{equation}
  \Upsilon_{\mu,\{a_i\}} = \frac{1}{L^3} \frac{\partial^2 F[\phi_i']}{\partial \delta_\mu^2},
\end{equation}
where $F$ is the free energy. Using the fundamental relations
\begin{equation}
  F = -T\ln Z, \quad Z = \mathrm{Tr}\,\mathrm{e}^{-\beta H},
\end{equation}
one can derive an expression for the second derivative of the free energy $F$ in terms of the first and second derivatives of the Hamiltonian $H$ with respect to the phase-twist parameter $\delta$:
\begin{equation}
  \frac{\partial^2 F}{\partial \delta^2} =
  \left\langle \frac{\partial^2 H}{\partial \delta^2} \right\rangle
  - \beta \left\langle \left(
      \frac{\partial H}{\partial\delta}
    \right)^2 \right\rangle.
  \label{FH}
\end{equation}
In deriving the above, we have used that $\langle \partial H / \partial \delta \rangle = 0$.

Looking at the Hamiltonian (\ref{lattf}) it is clear that any first derivative $\partial H / \partial \delta$ will be a linear combination of sine functions, and any second derivative $\partial H^2 / \partial^2 \delta$ will be a linear combination of cosine functions. In both cases, the coefficients will depend on the $a_i$. Because of the square in the second term in (\ref{FH}), the helicity modulus for a given linear combination of phases will depend not only on the corresponding single-phase helicity moduli, but also on certain cross terms that involve pairs of components.\cite{Dahl2008a, Herland2010} In our case the components are all equivalent, and consequently each of the single-phase helicity moduli are equal, as are each of the cross-term helicity moduli. Furthermore, because the phase sum couples to the vector potential, its helicity modulus is necessarily zero. Form this is follows that the ratio between the value of a single-phase helicity modulus and the value of a cross-term helicity modulus is given by simple combinatorics, and that all helicity moduli are proportional to each other (the phase-sum helicity modulus being the special case where the constant of proportionality is zero).

Because of the aforementioned proportionality of all helicity moduli, we could in principle equivalently use any helicity modulus (except that for the phase sum) to locate neutral transitions. We choose to use the average of all single-phase moduli and the negatives of all cross-term moduli, as this would apear to minimize statistical error. We denote this quantity, when rescaled to equal a single-phase modulus, simple as $\Upsilon$. In order to improve statistical precision, we determine a helicity modulus for a given temperature as as weighted average of the modulus determined from the data for that temperature and moduli obtained by reweighting from neighboring temperatures. We do this as follows: When reweighting from the nearest $n$ temperatures on either side, the unreweighted modulus has relative weight $n+1$ and the relative weights of the other moduli decrease by $1$ for each step away from the temperature to which reweigting is performed. Since the true value of the phase-sum helicity modulus is zero, the average over all temperatures of the squared phase-sum helicity modulus is a measure of the statistical error in the helicity moduli. The number $n$ is chosen to be the first natural number for which $n+1$ gives a larger error. However, we limit $n$ to a maximal value of $n = 8$ in order to limit the computational power required. Also, in some cases we measure helicity moduli in each coordinate direction and average them in order to further improve statistical precision. All this is motivated by the fact that for the systems studied here helicity moduli tend to have large statistical errors compared to other quantities that we study.

At the critical point of a continuous neutral transition, the helicity modulus is expected to scale as $1/L$, so that $L \Upsilon$ is a universal quantity. We use finite-size crossings of $L \Upsilon$, extrapolated to the thermodynamic limit, in order to locate neutral transitions.

\subsection{Results}

\begin{figure}
  \includegraphics{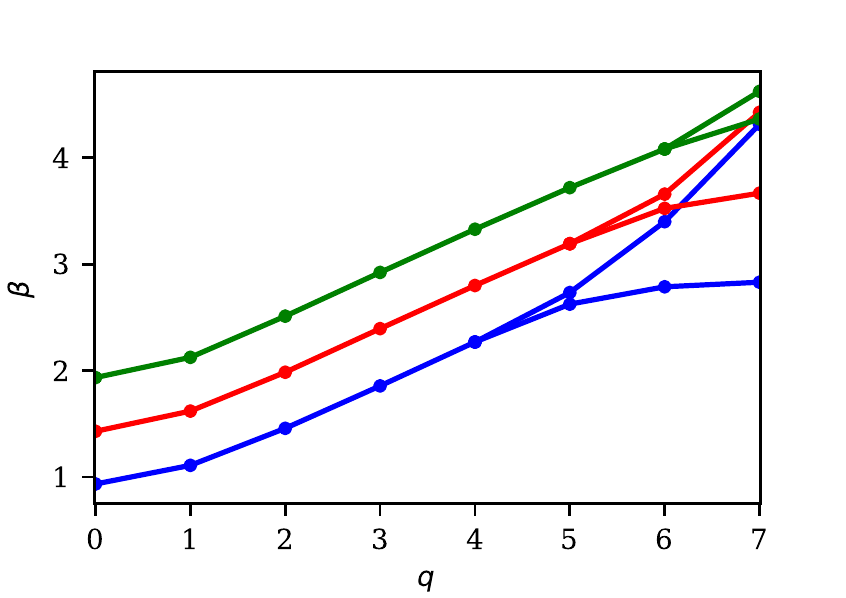}
  \caption{Phase diagrams for $N = 2$ (blue, lower diagram) $N = 3$ (red, middle diagram) and $N = 4$ (green, upper diagram). The phase diagrams show that for high enough electric charge $q$ there are new $\cp$-neutral phases  in which order is retained in the phase differences and relative-density degrees of freedom whilst superconducting order is absent. Errors are smaller than symbol sizes, and lines are a guide to the eye.}
  \label{phase_diag}
\end{figure}

Phase diagrams for the two-, three- and four-component cases are shown in Fig.~\ref{phase_diag}. Our main conclusion is that the systems have split transitions and phases with composite order for $N>2$. This occurs for high values of the coupling constant; if the temperature is increased the system first undergoes a phase transition where the Meissner effect disappears.

The remaining broken symmetry in the system is described by the following neutral terms in the original model:
\begin{multline}
  h_\mathrm{neutral} = \sum_{i,j>i} |\psi_i|^2 |\psi_j|^2 (\nabla \phi_{ij})^2 + \tfrac{1}{2} \sum_i \left( \nabla |\psi_i| \right)^2.
  \label{modes2}
\end{multline}
The new physics associated with these phases, in contrast to the $\mathrm{U}(1)^N$ and $\mathrm{SU}(2)$ cases, is that these phases cannot be mapped onto $\mathrm{O}(N)$ models. Instead these are $\cp$-neutral phases where there is order associated with the spontaneous breaking of symmetry only in the phase differences and relative densities between components.

For the highest charge simulated ($q = 7$) the shift in temperature for the superconducting phase transition to the completely ordered low-temperature state is much smaller than the corresponding shifts for the direct transitions at lower charges. The reason for this is that it is integer-flux excitations that govern the charged transitions, and increased charge makes these objects more tightly bound composites of fractional flux excitations [cf.\ the $U(1)^N$ case \cite{Babaev2002b, Smiseth2005, Sellin2016}].

\begin{figure}
	\includegraphics{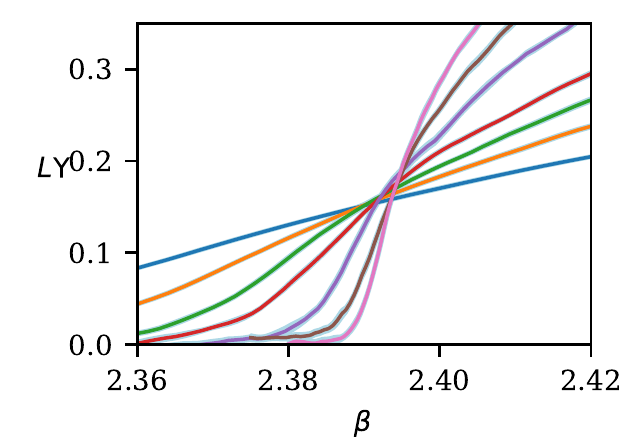}
	\includegraphics{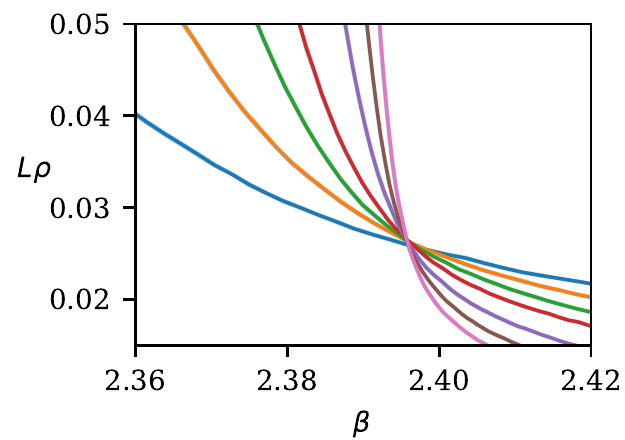}
	\includegraphics{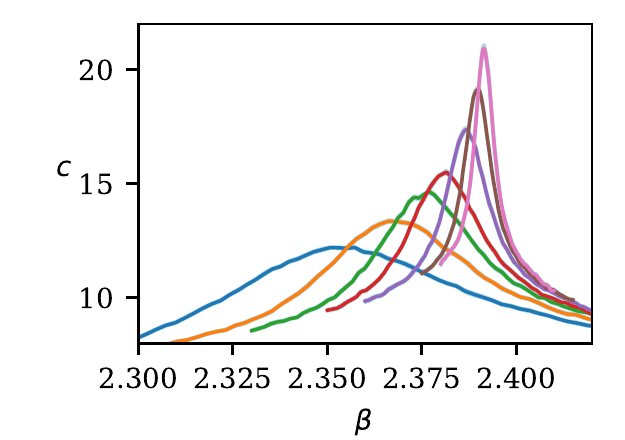}
  \caption{Finite-size crossings of $L\Upsilon$ (helicity modulus scaled by system size) versus inverse temperature $\beta$ (top) and of $L\rho$ (dual stiffness scaled by system size) versus inverse temperature (middle) as well as heat capacity $c = L^{-3}\,\mathrm{d}E/\mathrm{d}T$ versus inverse temperature (bottom) for $N = 3$, $q = 3$ and $L = 12,16,20,24,32,40,48$. Note that the scale for the inverse-temperature axis is different for the heat-capacity plot than for the other two plots; the crossings show much smaller finite-size corrections than the heat-capacity peaks. Errors are indicated by (narrow) shaded error regions.}
  \label{single_trans}
\end{figure}

\begin{figure}
	\includegraphics{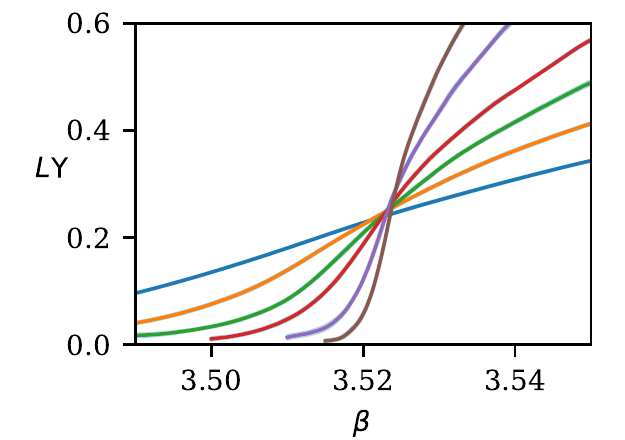}
	\includegraphics{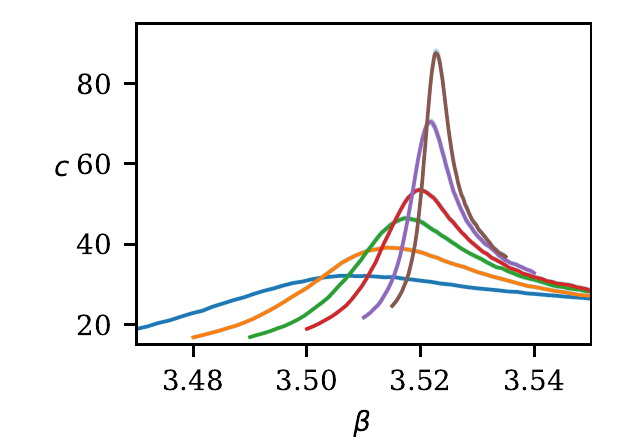}
  \caption{Finite-size crossings of $L\Upsilon$ (helicity modulus scaled by system size) versus inverse temperature $\beta$ (top) and heat capacity versus inverse temperature (bottom) for $N = 3$, $q = 6$ and $L = 12,16,20,24,32,40$. Errors are indicated by (narrow) shaded error regions.}
  \label{neutral_trans}
\end{figure}

\begin{figure}
	\includegraphics{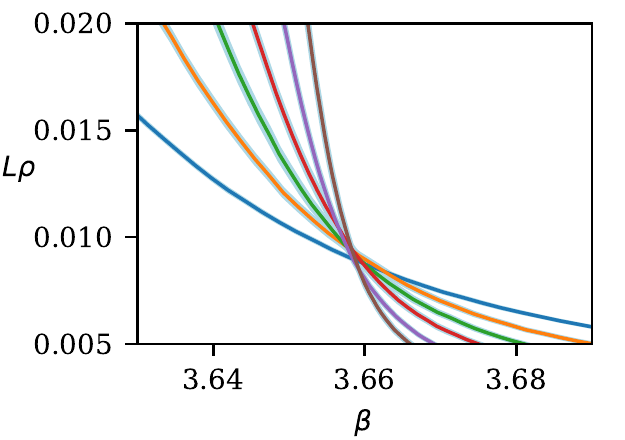}
	\includegraphics{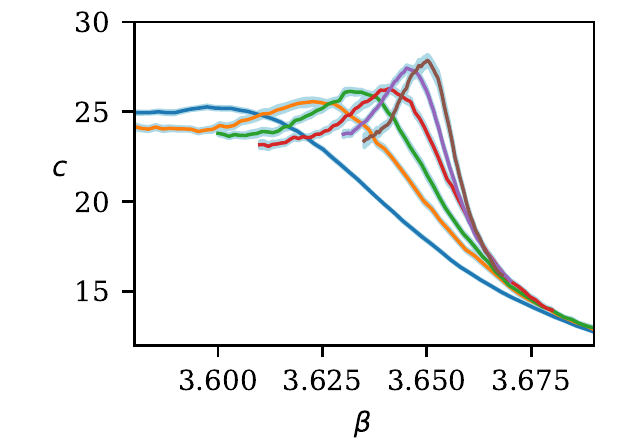}
  \caption{Finite-size crossings of $L\rho$ (dual stiffness scaled by system size) versus inverse temperature (top) and heat capacity versus inverse temperature (bottom) for $N = 3$, $q = 6$ and $L = 12,16,20,24,32,40$. Errors are indicated by shaded error regions.}
  \label{charged_trans}
\end{figure}

We now present more detailed results for some representative examples of the phase transitions in the aforementioned phase diagrams. We choose to consider the case $N = 3$, and the charges $q = 3$ for which there is a single transition and $q = 6$ for which there are two separate transitions. In Fig.~\ref{single_trans} we show the helicity moduli, dual stiffnesses and heat capacities for $q = 3$ and system sizes in the range $L = 12-48$. Within our numerical uncertainty, the two transitions occur at the same temperature. Also, there are clear peaks in heat capacity that increase with system size. In Fig.~\ref{neutral_trans} we show the helicity moduli and heat capacities for the neutral transition for $q = 6$ with system sizes in the range $L = 12-40$, and in Fig.~\ref{charged_trans} we show the heat capacities and dual stifnesses for the charged transition in systems with the same charge and sizes. The two transitions are clearly separated in temperature. Again, the transitions are associated with clear peaks in heat capacity that become more pronounced with increasing system size.

\section{Magnetic response of $SU(N)$ systems}

The $\cp$-composite neutral order demonstrated in the previous section is only possible because of the existence of integer-flux excitations. In a $\mathrm{U}(1)^N$ case these are vortices where each complex component has $2\pi$ winding around a shared core.\cite{Babaev2002b} As mentioned above, in contrast in the $\mathrm{SU}(N)$ case the integer flux vortices are not energetically stable. \cite{achucarro2000semilocal} In certain cases it is possible to stabilize these objects by breaking symmetry explicitly to $U(1)\times Z_2$; \cite{Garaud2013,Garaud2011} then vortices are characterized not only by an overall phase winding but also by a $\cp$ skyrmionic topological charge
when the cores of constituent fractional vortices do not coincide in space.

In this section we study the behavior of the models we consider in external magnetic field and at finite temperature. In particular we determine low-temperature vortex configurations. We find that vortices can be stabilized by the presence of external magnetic field.

This section consists of two parts: In the first part we describe how we implement external magnetic field and measure magnetic response numerically. In the second part we present and discuss our results involving external magnetic field.

\subsection{Numerical implementation}

For the purpose of implementing external magnetic field, we decompose the vector potential into a sum of two terms: $\mathbf{A}(\mathbf{r}) = \mathbf{A}_0(\mathbf{r}) + \mathbf{A}_1(\mathbf{r})$. Of these two terms, the first corresponds to a uniform magnetic field in the $z$ direction. This term, which is written in the Landau gauge, is held constant: $\mathbf{A}_0(\mathbf{r}) = (0, 2\pi x f, 0)$. The second term, $\mathbf{A}_1(\mathbf{r})$, is updated by Metropolis-Hastings updates and thus gives thermal fluctuations. Periodic boundary conditions are imposed on $\mathbf{A}_1(\mathbf{r})$, whence the contribution to total magnetic flux from this term is zero. Thus the total flux is equal to that given by the constant term $\mathbf{A}_0(\mathbf{r})$. Given that periodic boundary conditions are used, the value of $f$ must be such that $hqLf$ is an integer.

For the purpose of describing the response of the system to external magnetic field, we consider two quantities: vorticities and magnetic flux density. For both quantities, we average over the $z$ direction, which is the direction of applied field; this gives 2D images. In addition to these real-space images, we measure the absolute values of the Fourier transforms of these, i.e.\ structure factors. For all four types of quantity (direct images and structure factors of vorticities and magnetic field) we measure thermal averages. We remove the zero-wavevector component of the structure factors for clarity, and normalize the remaining components to the zero-wavevector component.

We now describe how vorticity is defined on a lattice. For a phase field defined on continuous space, it is clear how to determine if there is a vortex at a certain point: if so, the phase winds around this point. For a phase field defined only on the discrete points of a lattice, it is less clear what is means for there to be a vortex. The conventional way of counting the number of vortices on a given plaquette is this: The net vorticity of the plaquette is obtained by adding contributions from each of the links of the plaquette. For each link, consider the gauge-invariant phase difference $\chi_{i,k}(\mathbf{r})$. The contribution from a given link is the number of multiples of $2\pi$ that must be added to $\chi_{i,k}(\mathbf{r})$ in order to bring it into the primary interval $(-\pi, \pi]$; this can be positive or negative.

Note that on a continuum the circulation integral of the gradient of phase is quantized, and that this does not depend on the vector potential. However, for small enough integrals around the vortex core the inclusion of the vector potential makes negligible difference, and thus the concept of vorticity used here is consistent with the usual concept on a continuum.

In the limit of zero charge $q \rightarrow 0$ the magnetic field becomes completely uniform. We implement this numerically by choosing an initial condition for which the fluctuating part of the vector potential is zero, and then simply not updating the vector potential.

\subsection{Vortex structure in external field}

\begin{figure}
	\centerline{
		\includegraphics[scale=0.82]{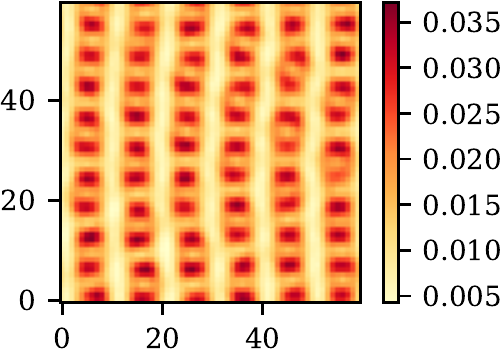}\hspace{4pt}
		\includegraphics[scale=0.82]{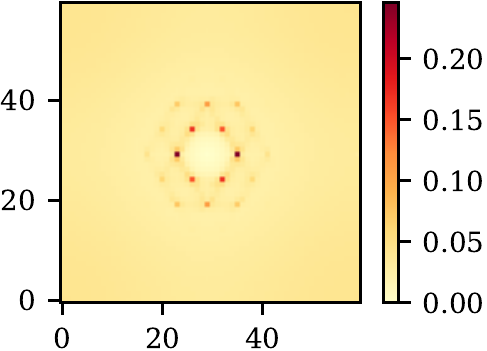} }
	\vspace{8pt}
	\centerline{
		\includegraphics[scale=0.82]{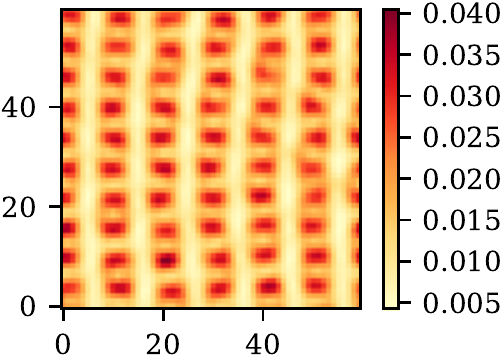}\hspace{4pt}
		\includegraphics[scale=0.82]{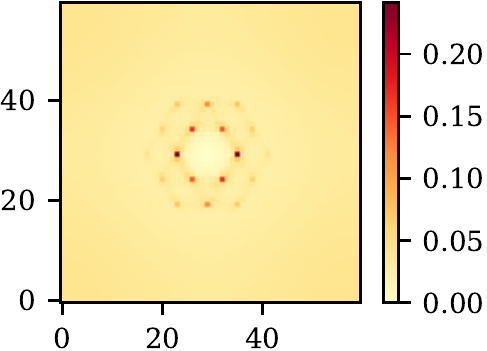} }
  \caption{Vortex patterns in external magnetic field in the two-component $q=0$ case. Both real-space thermal averages (left) and thermally averaged structure factors (right) are shown. The vortices form stripe patterns in each component, and the stripes of the two components are interlaced. The inverse temperature is $\beta = 1.5$. The external field is the weakest that can be applied in the Landau gauge for $L = 60$; thus there are a total of $60$ vortices in each component.}
  \label{vortex20-1}
\end{figure}

\begin{figure}
	\centerline{
		\includegraphics[scale=0.82]{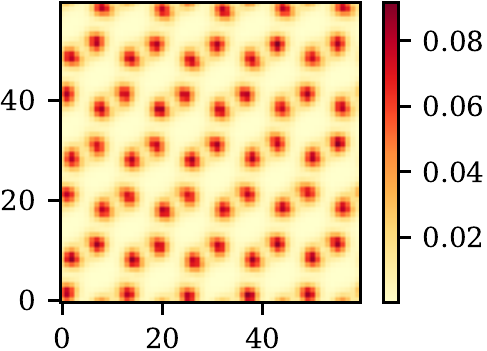}\hspace{4pt}
		\includegraphics[scale=0.82]{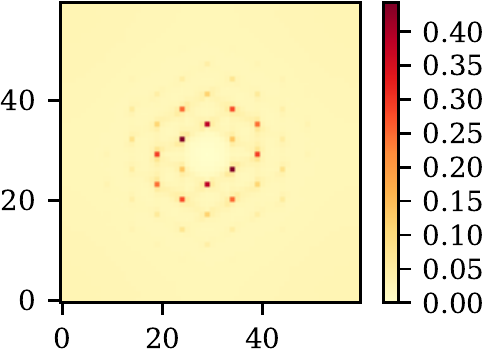} }
	\vspace{8pt}
	\centerline{
		\includegraphics[scale=0.82]{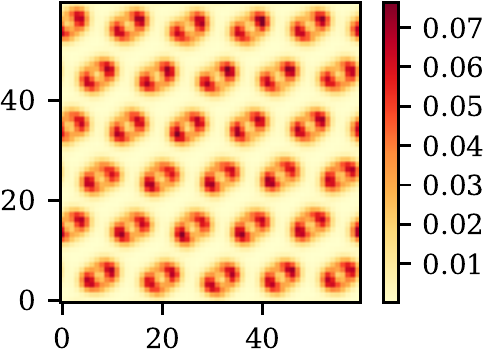}\hspace{4pt}
		\includegraphics[scale=0.82]{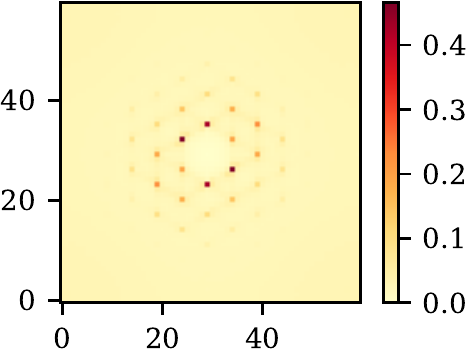}\hspace{4pt} }
  \caption{Same disaplayed quantities and parameters as in Fig.~\ref{vortex20-1}, except that here the inverse temperature $\beta = 2.5$. The vortices form patterns that are of different types than in Fig.~\ref{vortex20-1}, and which are not symmetric between the two components.}
  \label{vortex20-2}
\end{figure}

Vortex structures in neutral $\mathrm{SU}(2)$ systems under rotation
have been considered.\cite{Galteland2015} This regime is equivalent to the limit of infinite magnetic field penetration length or $q \to 0$ in the system under consideration. Vortex patterns in external magnetic field in the two-component zero-charge case are shown in Figs.~\ref{vortex20-1}-\ref{vortex20-2}. These vortex patterns are consistent with those found previously \cite{Galteland2015} in a model without a hard total-density constraint.

\begin{figure}
	\centerline{
		\includegraphics[scale=0.82]{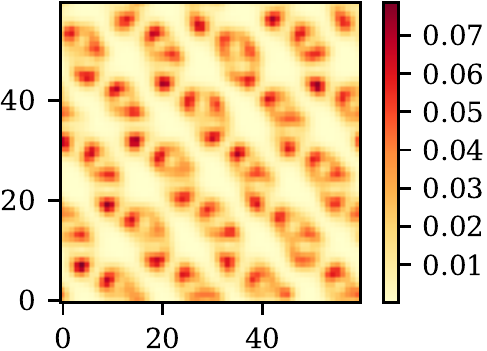}\hspace{4pt}
		\includegraphics[scale=0.82]{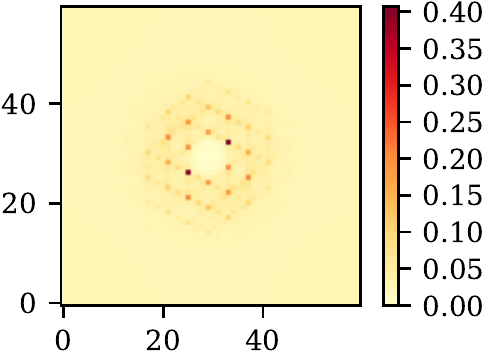} }
	\vspace{8pt}
	\centerline{
		\includegraphics[scale=0.82]{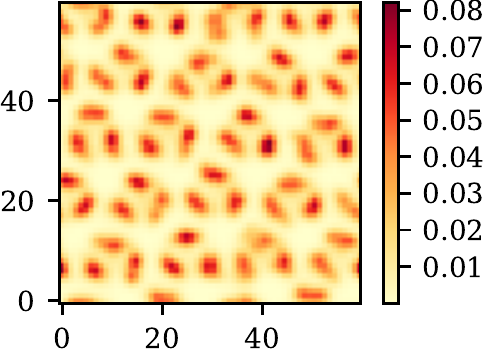}\hspace{4pt}
		\includegraphics[scale=0.82]{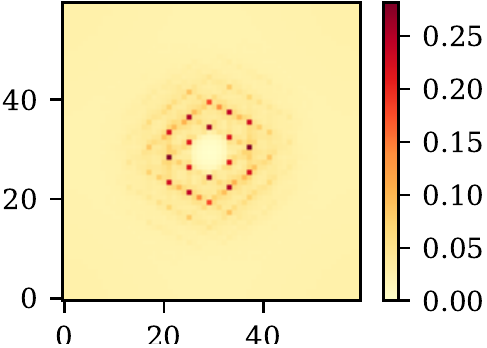} }
	\vspace{8pt}
	\centerline{
		\includegraphics[scale=0.82]{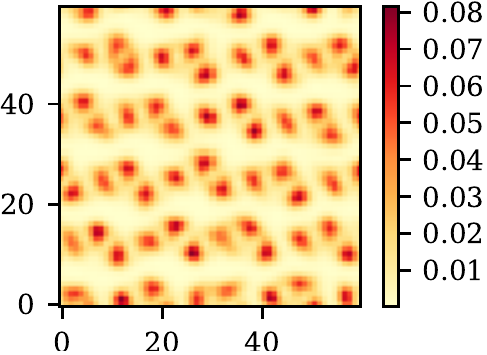}\hspace{4pt}
		\includegraphics[scale=0.82]{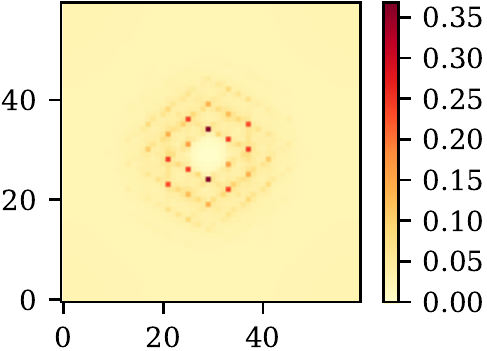} }
  \caption{Vortex patterns in external magnetic field in the three-component $q=0$ case. Both real-space thermal averages (left) and thermally averaged structure factors (right) are shown. The vortices form distinct patterns in each component. Each of the patterns is such that the vortices can naturally be divided into groups of three. The inverse temperature is $\beta = 5.5$. The external field is the weakest that can be applied in the Landau gauge for $L = 60$; thus there are a total of $60$ vortices in each component.}
  \label{vortex30}
\end{figure}

Fig.~\ref{vortex30} shows how these types of vortex state generalize to larger
numbers of components. More precisely, it displays a $q \to 0$ $\mathrm{SU}(3)$ superconductor in external field. In this three-component case the pattern is such that the vortices can be divided into groups of three. Such a pattern can be fully present only if the number of vortices is divisible by $3$. This is why we consider the system size $L = 60$ (rather than, say, $L = 64$).
Clearly, despite the instability of an individual vortex in the $\mathrm{SU}(3)$ case, under rotation, each component forms a regular vortex lattice, even in the presence of thermal fluctuations.

\begin{figure}
	\centerline{
		\includegraphics[scale=0.82]{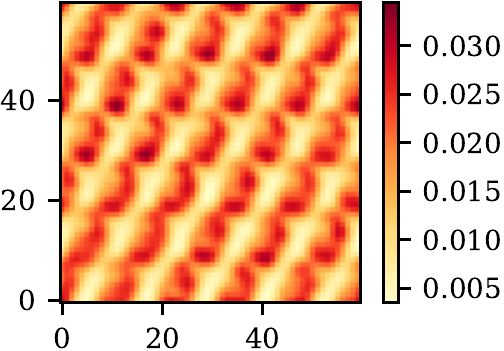}\hspace{4pt}
		\includegraphics[scale=0.82]{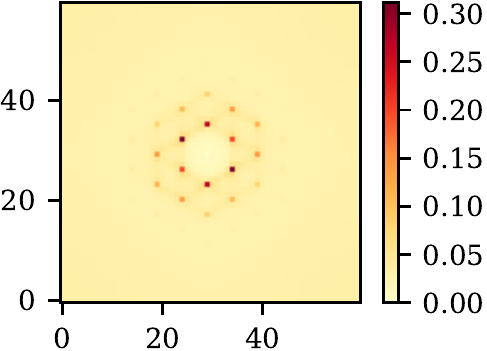}\hspace{4pt} }
	\vspace{8pt}
	\centerline{
		\includegraphics[scale=0.82]{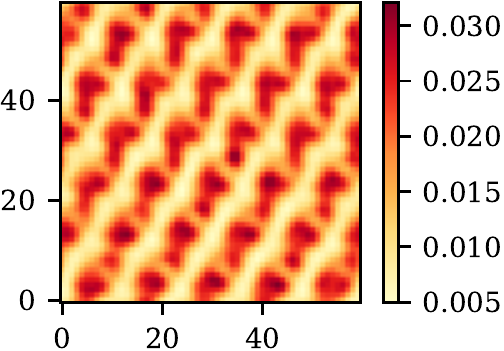}\hspace{4pt}
		\includegraphics[scale=0.82]{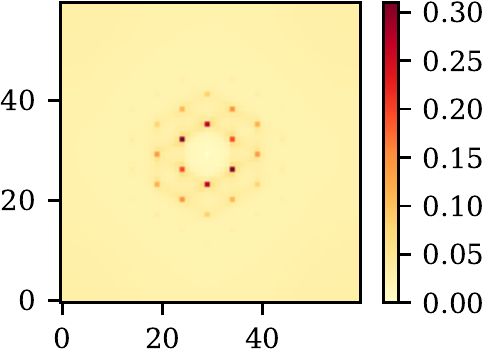}\hspace{4pt} }
	\vspace{8pt}
	\centerline{
		\includegraphics[scale=0.82]{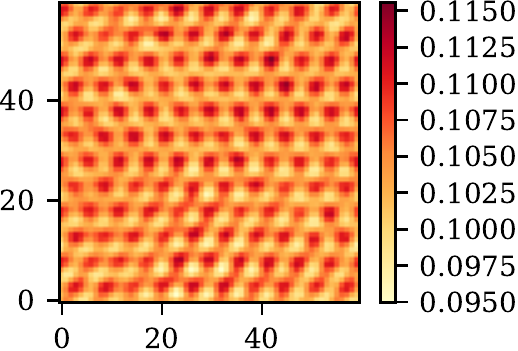}
		\includegraphics[scale=0.82]{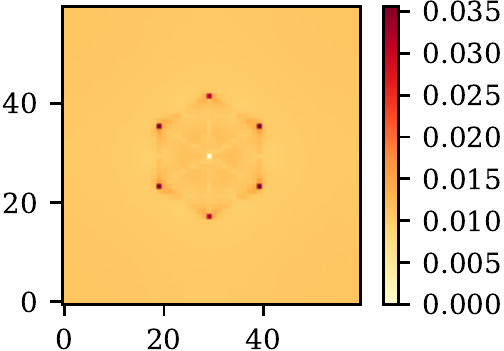} }
  \caption{Vortex patterns (top) and magnetic flux density (bottom) in external magnetic field in the two-component $q=1$ case. Both real-space thermal averages (left) and thermally averaged structure factors (right) are shown. The inverse temperature is $\beta = 2.5$. The external field is the weakest that can be applied in the Landau gauge for $L = 60$; thus there are a total of $60$ vortices in each component.}
  \label{vortex21-1}
\end{figure}

\begin{figure}
	\centerline{
		\includegraphics[scale=0.82]{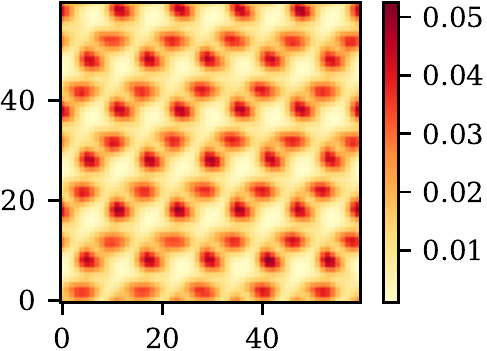}\hspace{4pt}
		\includegraphics[scale=0.82]{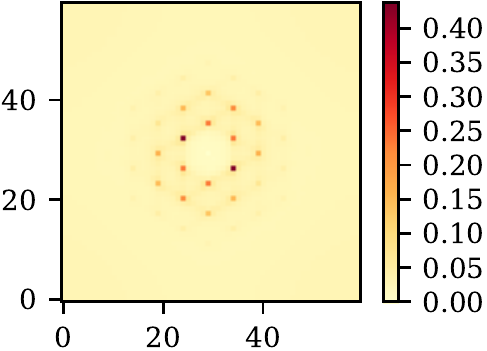}\hspace{4pt} }
	\vspace{8pt}
	\centerline{
		\includegraphics[scale=0.82]{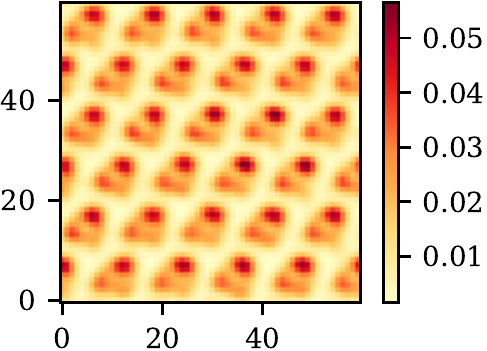}\hspace{4pt}
		\includegraphics[scale=0.82]{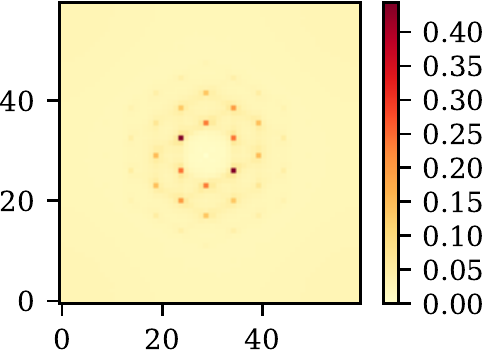}\hspace{4pt} }
	\vspace{8pt}
	\centerline{
		\includegraphics[scale=0.82]{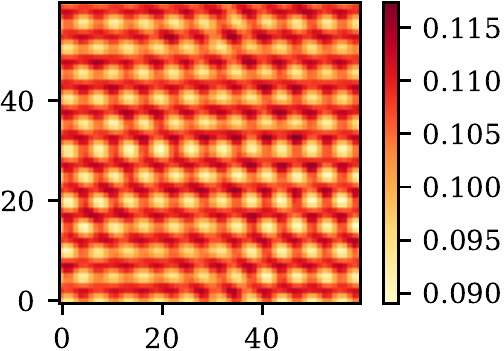}
		\includegraphics[scale=0.82]{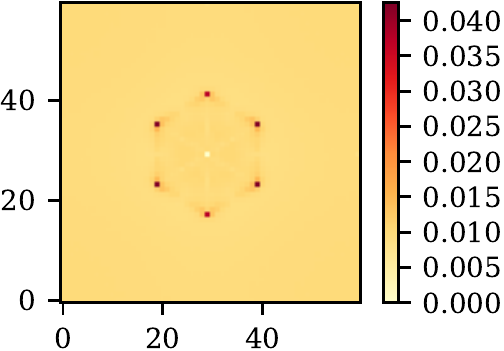} }
  \caption{Same disaplayed quantities and parameters as in Fig.~\ref{vortex21-1}, except that here the inverse temperature $\beta = 3.0$.}
  \label{vortex21-2}
\end{figure}

\begin{figure}
	\centerline{
		\includegraphics[scale=0.82]{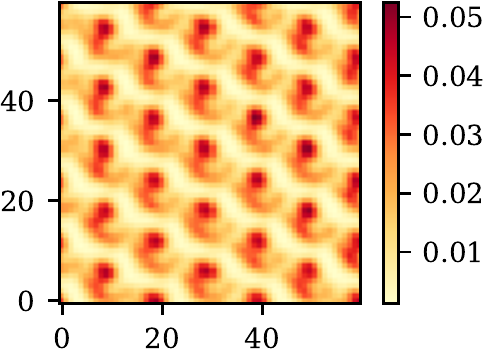}\hspace{4pt}
		\includegraphics[scale=0.82]{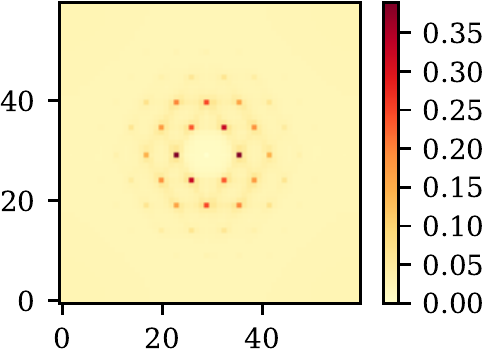} }
	\vspace{8pt}
	\centerline{
		\includegraphics[scale=0.82]{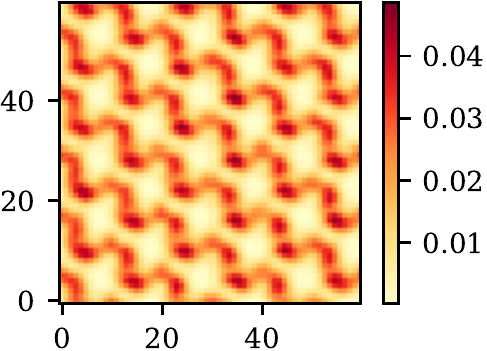}\hspace{4pt}
		\includegraphics[scale=0.82]{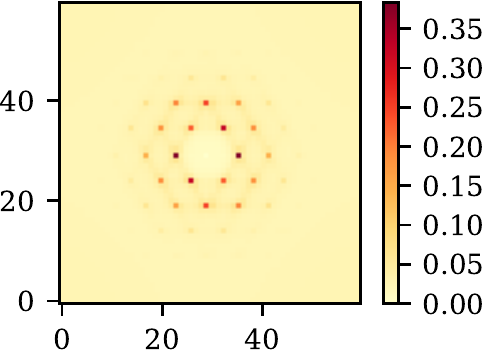} }
	\vspace{8pt}
	\centerline{
		\includegraphics[scale=0.82]{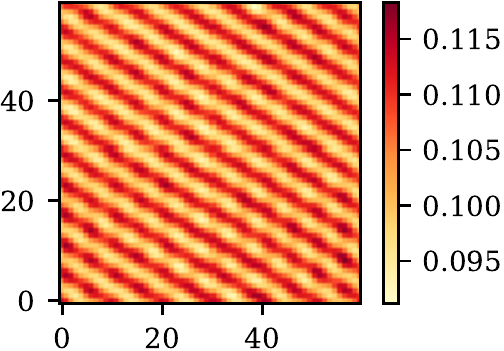}
		\includegraphics[scale=0.82]{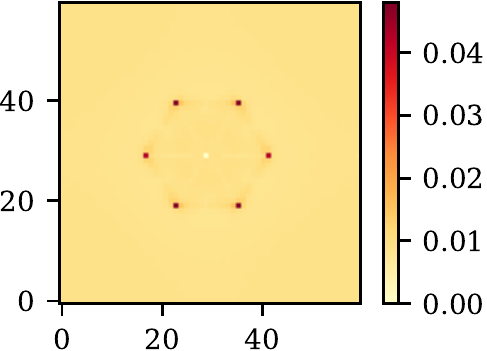} }
  \caption{Same disaplayed quantities and parameters as in Fig.~\ref{vortex21-1}, except that here the inverse temperature $\beta = 3.5$.}
  \label{vortex21-3}
\end{figure}

Next we move to the effect of finite screening lengths. Vortex patterns in external magnetic field in the case of two components and finite screening length ($q = 1$) are shown in Figs.~\ref{vortex21-1}-\ref{vortex21-3}. Since the charge is nonzero the magnetic field is nonuniform and is also shown. The vorticities, as defined above, form distinctive patterns. The magnetic field also forms patterns that are not necessarily similar to triangular lattices formed by Abrikosov lattices. The structure factors for the magnetic field in each case have exactly six peaks. However, these six peaks are not necessarily of equal strength. (Recall that we set the zero-wavevector component of the structure factors to zero for clarity.) In contrast to a zero-temperature study in finite geometry,\cite{Garaud2014} we find that magnetic field forms a well defined hexagonal lattice. Each peak of magnetic field corresponds to half a flux quantum. The regular structure that we obtain demonstrates a lattice of half-quantum vortices which is stable at finite temperature. Remarkably, the half-quantum vortices are non-topological excitations in the $\mathrm{SU}(N)$ models, but that lattice can be interpreted also as a lattice of composite  integer flux objects where the flux is split into two half-quanta vortices.

\section{Conclusion}
$\mathrm{SU}(N)$ theories with non-compact Abelian gauge fields are of interest as effective theories of various condensed-matter systems; however, the most investigated case is that of $\mathrm{SU}(2)$. We have considered $\mathrm{SU}(N)$-symmetric Ginzburg-Landau models at finite temperature with a focus on the case $N>2$. We found that non-topological vortices play important roles in these systems. First we pointed out that at sufficiently large coupling constants the system has states with composite order which cannot be mapped onto $S^1$ or $S^2$ neutral systems like the previously discussed paired phases in $\mathrm{U}(1)\times \mathrm{U}(1)$ and $\mathrm{SU}(2)$ theories. The new phases display no Meissner effect but have a spontaneously broken symmetry associated with the relative phases and relative densities of the components. We term these phases $\cp$-neutral states. These states result from proliferation of non-topological vortices that eliminate the Meissner effect, yet cannot disorder the $\cp$-neutral sector of the model. Given the importance of these non-topological vortices, we study these systems in external field and at finite temperature. These systems are not superconductors or superfluids in the usual sense because of the lack of a $U(1)$ topological invariant. Nonetheless we demonstrated that, even at finite temperature, well defined  lattices of non-topological vortices are formed in externally applied magnetic field.

\begin{acknowledgments}
The work was supported by the Swedish Research Council Grant No.\ 642-2013-7837 and by the G\"oran Gustafsson Foundation for Research in Natural Sciences and Medicine. The simulations were performed on resources provided by the Swedish National Infrastructure for Computing (SNIC) at the National Supercomputer Center at Link\"oping, Sweden.
\end{acknowledgments}

%\bibliography{weston}
%merlin.mbs apsrev4-1.bst 2010-07-25 4.21a (PWD, AO, DPC) hacked
%Control: key (0)
%Control: author (8) initials jnrlst
%Control: editor formatted (1) identically to author
%Control: production of article title (-1) disabled
%Control: page (0) single
%Control: year (1) truncated
%Control: production of eprint (0) enabled
%

\end{document}